\documentclass[%
 reprint,
 amsmath,amssymb,
 aps,
prl,
]{revtex4-2}

\usepackage{graphicx}
\usepackage{dcolumn}
\usepackage{bm}

\begin{document}

\title{Electrostatic trapping of $\mathrm{N_{2}}$ molecules in high Rydberg states}

\author{M. H. Rayment}
\affiliation{Department of Physics and Astronomy, University College London, Gower Street, London WC1E 6BT, United Kingdom}
\author{S. D. Hogan}
\affiliation{Department of Physics and Astronomy, University College London, Gower Street, London WC1E 6BT, United Kingdom}
\email{s.hogan@ucl.ac.uk}

\date{\today}

\begin{abstract}
N$_2$ molecules traveling in pulsed supersonic beams have been excited from their ${\mathrm{X\,^1\Sigma_{\mathrm{g}}^+}}$ ground electronic state to long-lived Rydberg states with principal quantum numbers between 39 and 48 using a resonance-enhanced two-color three-photon excitation scheme. The Rydberg states populated had static electric dipole moments exceeding $5000$~D which allowed deceleration of the molecules to rest in the laboratory-fixed frame of reference and three-dimensional trapping using inhomogeneous electric fields. The trapped molecules were confined for up to 10~ms, with effective trap decay time constants increasing with principal quantum number, and ranging from $450~\mu$s to $700~\mu$s. These observations, and comparison with the results of similar measurements with He atoms, indicate that the decay dynamics of the trapped Rydberg N$_2$ molecules are dominated by spontaneous emission and do not exhibit significant contributions from effects of intramolecular interactions that lead to non-radiative decay. 
\end{abstract}

\maketitle

The homonuclear diatomic molecules H$_2$, N$_2$ and O$_2$ are central to the physics and chemistry of the Earth's atmosphere and play important roles in tests of molecular quantum mechanics~\cite{Holsch2019,Beyer2019DeterminationH2,Sprengers2003,Salumbides2009}, searches for time variations of fundamental constants~\cite{Reinhold2006IndicationSpectra,Ubachs2007OnLaboratory,kajita14a}, and studies of quantum mechanical effects in low temperature reactions~\cite{Allmendinger2016NewReaction,Allmendinger2016ObservationEnergies,Hoveler2022ObservationReaction,margulis23a}. For many quantum-state-resolved experiments with these species, it is desirable to prepare translationally cold samples in velocity-controlled beams or traps. Since O$_2$ is paramagnetic in its $\mathrm{X\,^3\Sigma_{\mathrm{g}}^-}$ ground state, deceleration and trapping can be achieved using inhomogeneous magnetic fields~\cite{Wiederkehr2012Velocity-tunableDeceleration,Akerman2015SimultaneousBeam,Segev2019CollisionsTrap}. H$_2$ and N$_2$ in their ${\mathrm{X\,^1\Sigma_{\mathrm{g}}}}$ ground states do not exhibit significant magnetic or electric dipole moments. Consequently, forces cannot easily be exerted on them using magnetic or electric field gradients. However, if these molecules are excited to Rydberg states with static electric dipole moments $\gtrsim1000$~D, deceleration and trapping is possible using inhomogeneous electric fields though the methods of Rydberg-Stark deceleration~\cite{Procter2003,Yamakita2004,Hogan2009,Seiler2011a,SeilerPhDthesis,Hogan2016}. Cold Rydberg H$_2$, and NO molecules trapped or decelerated in this way have enabled studies of slow excited-state decay processes~\cite{Hogan2009,Seiler2011a,SeilerPhDthesis,Deller2020PRL,Rayment2021,Rayment2022}, and ion-molecule reactions at low temperature~\cite{Allmendinger2016NewReaction,Allmendinger2016ObservationEnergies,Hoveler2022ObservationReaction}. However, challenges in identifying laser photoexcitation schemes to prepare sufficiently long-lived Rydberg states, have meant that up to now these techniques were not applied in experiments with heavier homonuclear molecules.

Here, we prepare high Rydberg states in N$_2$ with lifetimes $>100~\mu$s using a $(2+1')$ resonance-enhanced two-color three-photon excitation scheme. This, together with subsequent $\ell$- and $M_{N}$-mixing ($\ell$ and $M_{N}$ are the electron orbital angular momentum quantum number and the azimuthal quantum number associated with the total angular momentum excluding spin, $\vec{N}$, of the molecule, respectively) in weak electric fields close to the time of photoexcitation led to the population of long-lived $\ell$-mixed Rydberg-Stark states with large static electric dipole moments. Molecules in the subset of these states with positive Stark shifts, i.e., the low-field-seeking (LFS) states, were selectively decelerated and electrostatically trapped. 

\begin{figure}[t]
\centering
    \includegraphics[width = \columnwidth]{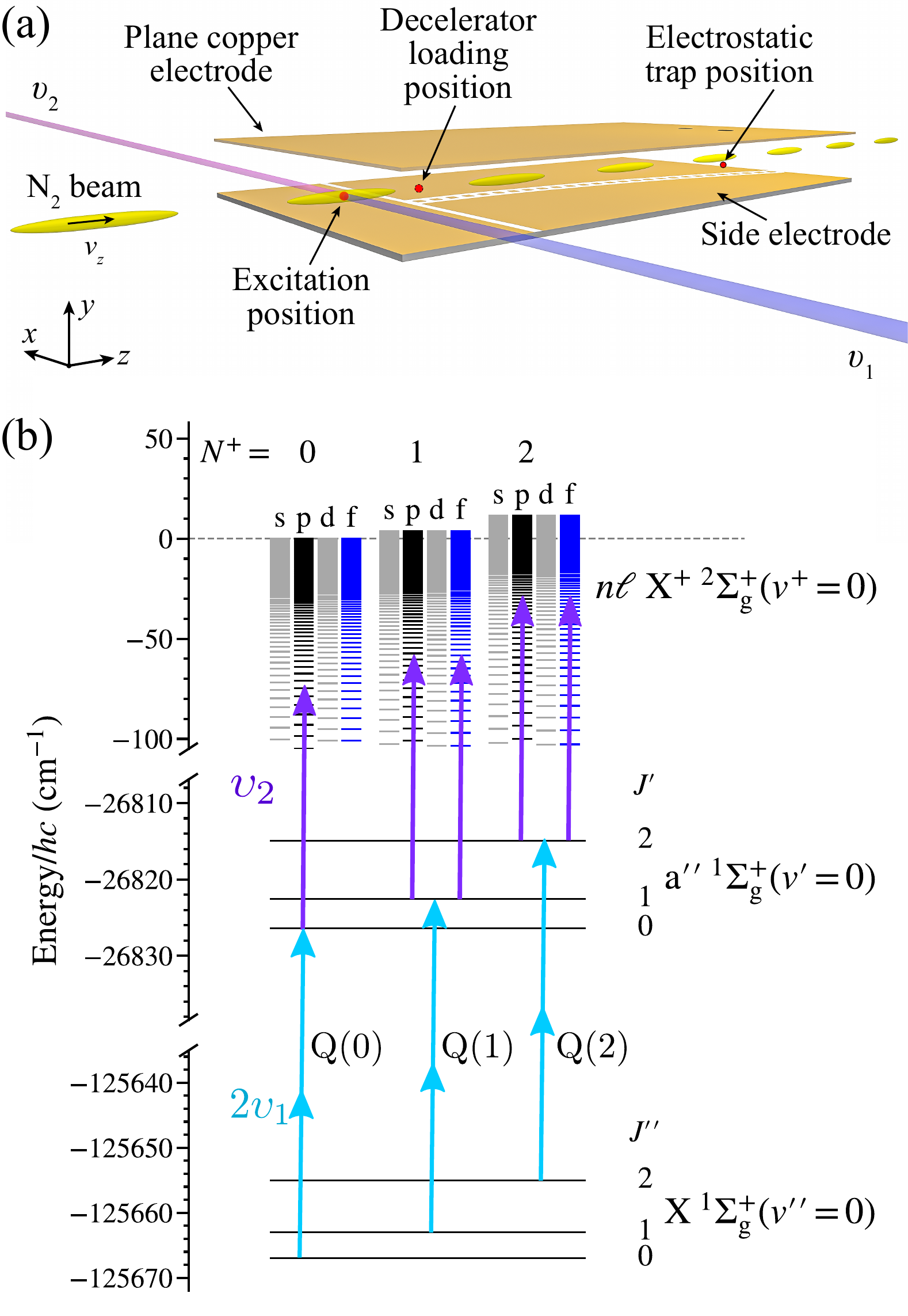}
    \caption{(a) Schematic diagram of Rydberg-Stark decelerator. (b) Resonance-enhanced two-color three-photon excitation scheme used to populate long-lived Rydberg states in N$_2$.}
    \label{Fig1}
\end{figure}

The Rydberg-Stark decelerator used here is depicted in Fig.~\ref{Fig1}(a) and was identical to that described previously~\cite{Deller2020PRL,Rayment2021}. Pure pulsed supersonic beams of N$_2$ were generated with mean longitudinal speeds $\bar{v}_{z} = 810~\mathrm{m\,s}^{-1}$. After collimation, the beams entered the decelerator, which was enclosed in a copper heat shield and cooled to $30$~K to minimize effects of blackbody radiation. Between a pair of parallel electrodes within the decelerator structure, the molecules were photoexcited to selected Rydberg states using the $(2+1')$ resonance enhanced ${\mathrm{X}\,\mathrm{^{1}\Sigma^{+}_{g}} \left( v^{\prime \prime}=0, \, J^{\prime \prime}  \right)} \rightarrow {\mathrm{a^{\prime \prime}}\,\mathrm{^{1}\Sigma^{+}_{g}} \left( v^{\prime}=0 , \, J^{\prime}  \right)} \rightarrow {n\ell\, \mathrm{X}^{+} \, \mathrm{^{2}\Sigma^{+}_{g}}\left(v^{+}=0, \, N^{+} \right)}$ excitation scheme in Fig.~\ref{Fig1}(b) ($v$ and $J$ are vibrational and rotational quantum numbers, respectively)~\cite{Merkt1995RotationalNitrogen,Mackenzie1995RotationalState}. This was driven using the frequency-tripled and fundamental outputs of two pulsed dye lasers. The first, two-photon Q-branch transition was driven at $\upsilon_{1} = 49420.5~\mathrm{cm^{-1}}$ [${\sim 150~\mathrm{\mu J/pulse}}$; $\sim100~\mu\mathrm{m}$ full width at half maximum (FWHM)]. Transitions from the $\mathrm{a^{\prime\prime}}$ state to high Rydberg states were driven using the second counter-propagating beam at $\upsilon_{2} = 26735 - 26810~\mathrm{cm}^{-1}$ [$\sim 4~\mathrm{mJ/pulse}$; $\sim2$~mm FWHM]. This resulted predominantly in the excitation of Rydberg states with ${n\mathrm{p}(N^{+})}$ or ${n\mathrm{f}(N^{+})}$ character for which $N^{+} = J^{\prime} = J^{\prime\prime}$~\cite{Mackenzie1995RotationalState}, and ensembles of excited molecules with rotational temperatures reflecting those of the ground-state supersonic beams ($\sim10$~K). 

After excitation, the molecules traveled $4.8$~mm to where they were loaded into a single traveling electric trap of the Rydberg-Stark decelerator~\cite{Lancuba2014}. This comprised a curved array (518.5~mm radius of curvature) of 0.5~mm$\times$0.5~mm electrodes with a 1-mm center-to-center spacing, and was operated with 5 sinusoidally oscillating potentials of amplitude $V_{0} = 149~\mathrm{V}$. A potential $-V_{0}/2$ was applied to the parallel plane copper electrode, separated from the array by 3~mm, during decelerator operation. The potentials on the side electrodes (ground planes of the transmission line) were maintained at 0~V for photoexcitation, deceleration and trapping. \emph{In situ} detection of the Rydberg molecules was achieved by pulsed electric field ionization (PFI)~\cite{Lancuba2016}. To implement this, the decelerator potentials were reduced to zero and pulsed potentials of $+460~\mathrm{V}$ were applied to the side electrodes. This allowed N$_2^+$ ions generated by PFI at the electrostatic trapping position, $\sim105$~mm from the excitation position, or a second location on the molecular-beam axis a similar distance from the excitation region, to be accelerated though one of two 2-mm-diameter apertures in the plane copper electrode to a microchannel plate detector.

\begin{figure}[tb]
\centering
    \includegraphics[width = \columnwidth]{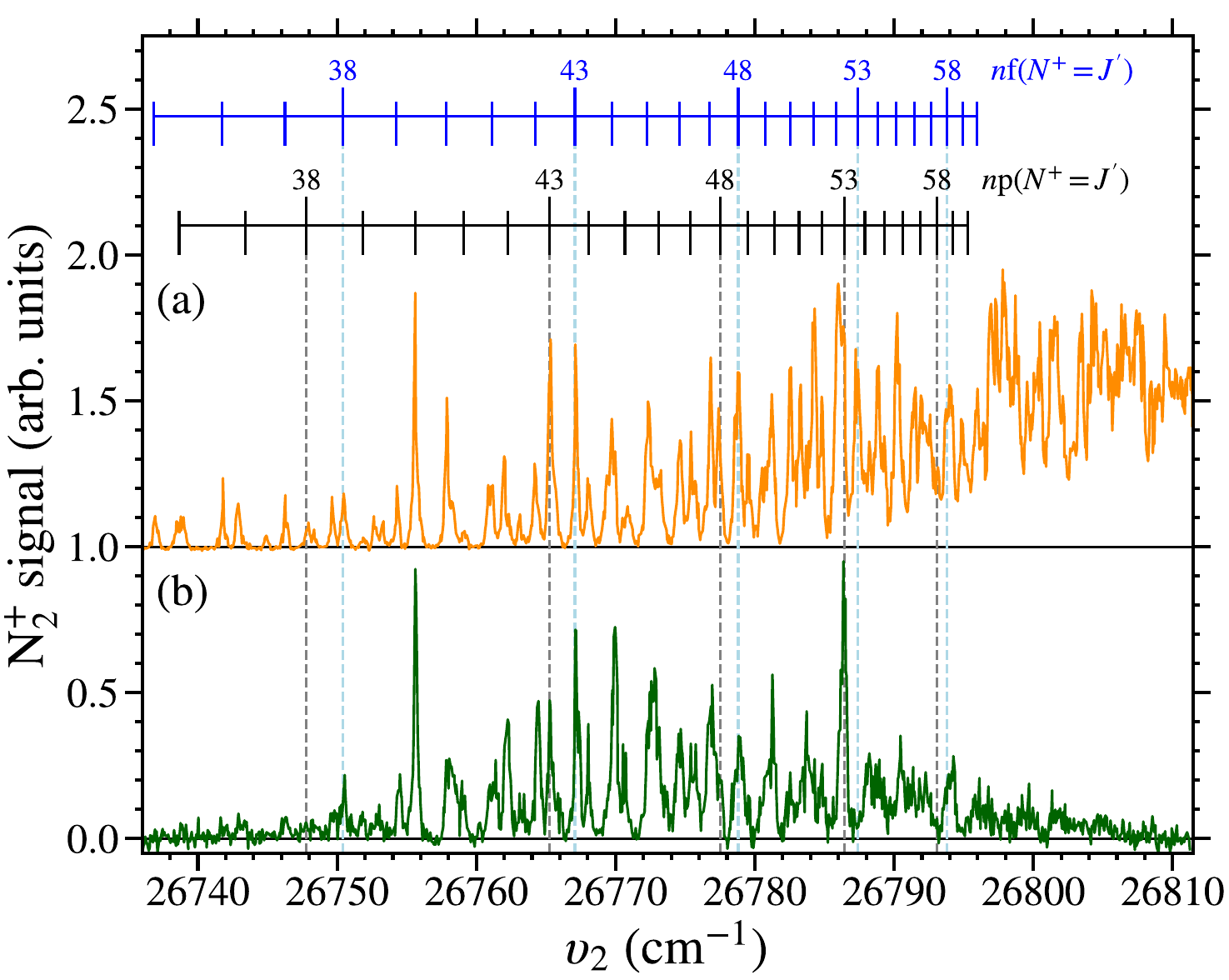}
    \caption{Laser photoexcitation spectra recorded by delayed PFI with (a)~the decelerator off and detection on the molecular beam axis at $t_{\mathrm{PFI}}=130~\mathrm{\mu s}$, and (b)~following deceleration and electrostatic trapping for $t_{\mathrm{trap}}=100~\mathrm{\mu s}$.}
    \label{Fig2}
\end{figure}

The range of values of $n$ of the accessible long-lived Rydberg states was identified by recording laser photoexcitation spectra with detection after a time $t_{\mathrm{PFI}} = 130~\mathrm{\mu s}$ as shown in Fig.~\ref{Fig2}(a). In this case, PFI was performed with ion extraction through the detection aperture on the molecular beam axis. This spectrum exhibits a high density of features, but in general Rydberg states with sufficient lifetimes for detection were populated by excitation on ${n\mathrm{f}(N^{+}=J^{\prime})}$ resonances when $n > 35$. The small quantum defects ($\delta_{n\mathrm{f}}\simeq0.01$~\cite{Wilson1993IntensityN2}), of the ${n\mathrm{f}(N^{+})}$ states in N$_2$, meant that $\ell$- and $M_N$-mixing induced by weak time-varying electric fields -- associated with laboratory noise, and arising from collisions with charged particles -- close to the photoexcitation time led to the evolution of some excited-state population into longer-lived ‘hydrogenic' Rydberg-Stark states~\cite{Chupka1993,Merkt1994OnIonization}. A subset of these states with $|M_{N}|\geq4$ have predominantly high-$\ell$, i.e., $\ell>3$, character and their rapid predissociation is inhibited. Molecules in these long-lived states were selectively detected in Fig.~\ref{Fig2}(a), and suitable for deceleration and trapping. Short-lived, predissociative ${n\mathrm{p}(N^{+})}$ states, with quantum defects of $\delta_{n\mathrm{p}}\simeq0.646$~\cite{Merkt1992RotationallyNitrogen}, are also accessible using the excitation scheme in Fig.~\ref{Fig1}(b) but require larger electric fields for complete $\ell$-mixing. Excitation on $n\mathrm{p}(N^{+})$ resonances therefore only led to the population of long-lived Rydberg-Stark states if intramolecular charge-multipole interactions with accidentally near degenerate states with smaller quantum defects occurred~\cite{Bixon1996a,Bixon1996}. 

\begin{figure}[tb]
\centering
    \includegraphics[width = \columnwidth]{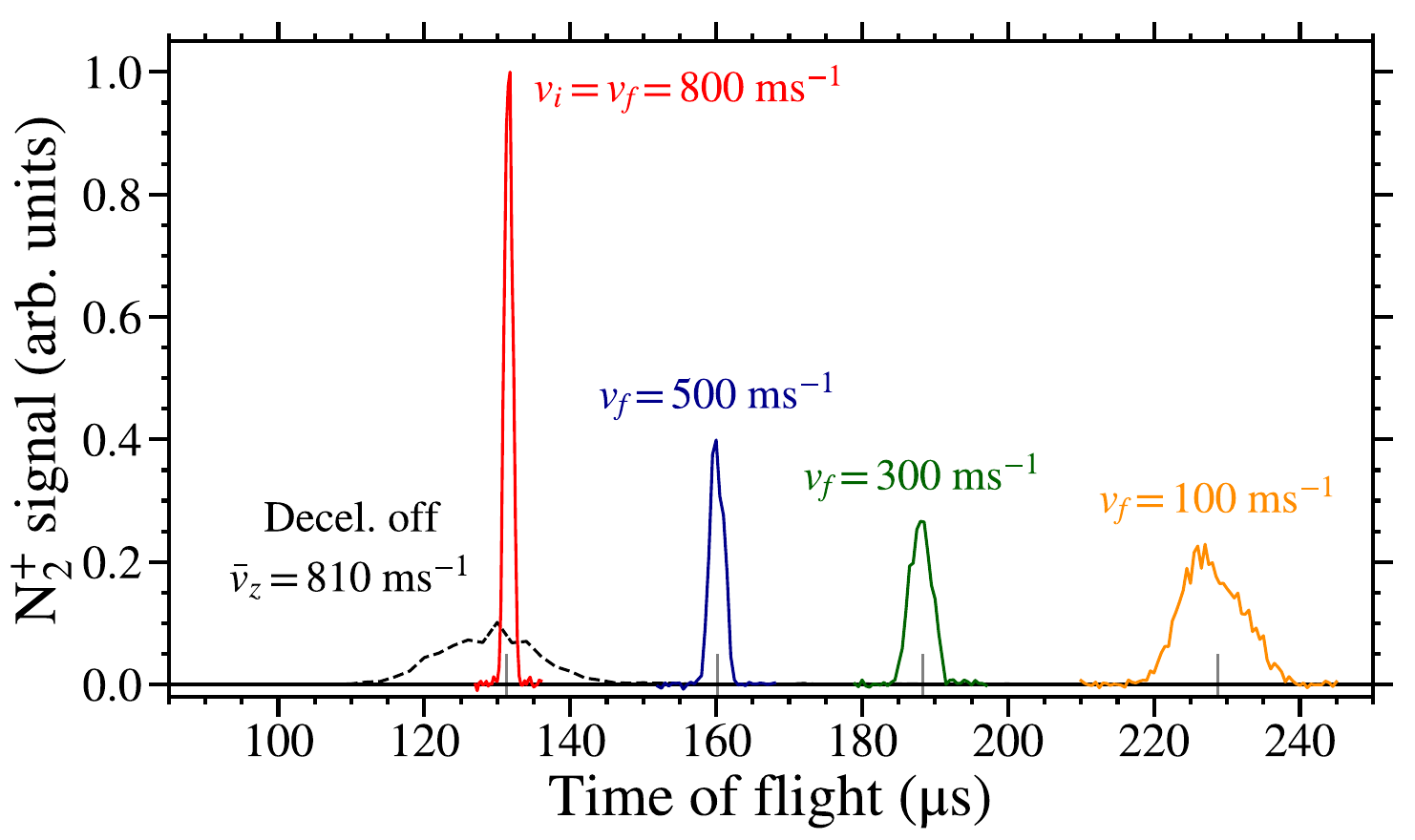}
    \caption{TOF distributions of Rydberg $\mathrm{N_{2}}$ molecules excited on the ${43\mathrm{f}(N^{+}=J^{\prime})}$ resonance with $\upsilon_{2} = 26767.10~\mathrm{cm^{-1}}$. Data recorded with the decelerator off are indicated by the dashed black curve with a maximum close to 130~$\mu$s. Measurements made with the decelerator on to guide the molecules at 800~m\,s$^{-1}$ (red curve) and decelerate them from $v_{\mathrm{i}} = 800$~m\,s$^{-1}$ to $v_{\mathrm{f}}=500$, 300 and 100~m\,s$^{-1}$ are also shown.}
    \label{Fig3}
\end{figure}

To guide and decelerate the Rydberg N$_2$ molecules, $\upsilon_2$ was set to $26767.10~\mathrm{cm^{-1}}$, i.e., resonant with the transitions to the ${43\mathrm{f}(N^{+}=J^{\prime})}$ states. This resulted in the population of long-lived hydrogenic Rydberg-Stark states with $n=43$ and $N^{+} = J^{\prime}$, denoted $n(N^+) = 43(N^+)$. The mean longitudinal speed, $\overline{v}_z = 810$~m\,s$^{-1}$ (standard deviation $\sigma_{v_z} \simeq 40$~m\,s$^{-1}$) of the excited molecules was determined by measuring their time of flight (TOF) from the excitation position to the on-axis detection position, as shown in Fig.~\ref{Fig3} (dashed black curve).

The molecules were loaded into a single traveling electric trap of the decelerator by activating the oscillating potentials 6~$\mu$s after photoexcitation. A TOF distribution recorded with the decelerator operated to guide the molecules at 800~m\,s$^{-1}$ to the off-axis detection position is shown in Fig.~\ref{Fig3} (tall red curve). Although Rydberg-Stark states with positive and negative Stark shifts were populated upon excitation~\cite{Deller2020}, the decelerator guided, and hence filtered, only the subset of molecules in the outer LFS states. The reduced FWHM of the TOF distribution of the guided molecules is a consequence of their localization about one single traveling electric field minimum when transported in the decelerator, such that they do not disperse and all pass the detection region simultaneously. 

To decelerate the molecules, the frequency of the oscillating potentials was chirped to slow the trap -- and the molecules confined within it -- from an initial speed $v_{\mathrm{i}}=800~\mathrm{m \, s^{-1}}$ to final speeds of $v_{\mathrm{f}} = 500$, $300$ and $100~\mathrm{m \, s^{-1}}$ with tangential accelerations $a_{\mathrm{t}} = -1.95 \times 10^{6}$, $-2.74 \times 10^{6}$ and $-3.14 \times 10^{6}~\mathrm{m \, s^{-2}}$, respectively. The effect of this on the TOF distributions is seen in Fig.~\ref{Fig3} where the vertical bars beneath each dataset indicate the flight time calculated from the classical equations of motion. The increase in the FWHM of the TOF distributions as $v_{\mathrm{f}}$ decreases, reflects the longer time taken by the slower molecules to pass the detection position~\cite{Deller2020PRL}. The reduction in the $\mathrm{N_{2}^{+}}$ signal as $\left| a_{\mathrm{t}} \right|$ increases, reflects the reduced phase-space acceptance of the decelerator at higher accelerations, and the decay of the molecules.

\begin{figure}[tb]
\centering
    \includegraphics[width = \columnwidth]{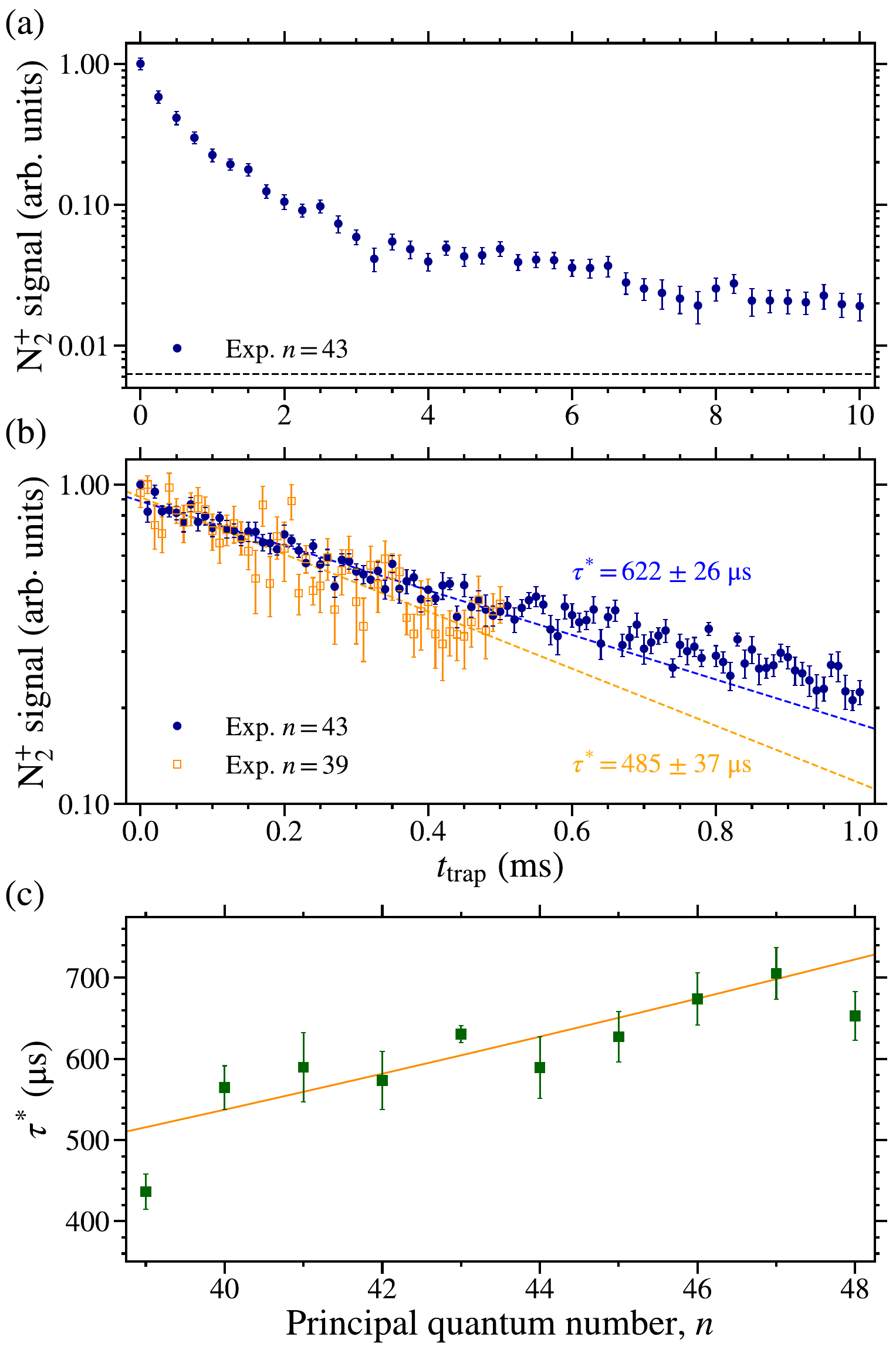}
    \caption{Decay of Rydberg $\mathrm{N_{2}}$ molecules from the electrostatic trap for trapping times up to (a)~$10~\mathrm{ms}$, and (b)~$1~\mathrm{ms}$. The black dashed line in (a) represents the standard deviation of the background signal. The molecules were photoexcited on the $43\mathrm{f}(N^{+}=J^{\prime})$ resonance ($\upsilon_{2} = 26767.10~\mathrm{cm^{-1}}$) in (a) and (b), or the $39\mathrm{f}(N^{+}=J^{\prime})$ resonance ($\upsilon_{2} = 26754.21~\mathrm{cm^{-1}}$) in (b). In (b) single exponential functions fit to the data between $t_{\mathrm{trap}}=50$ and $500~\mathrm{\mu s}$ are indicated by the dashed curves. (c)~Effective trap decay time constants, $\tau^*$, for molecules excited on ${n\mathrm{f}(N^{+}=J^{\prime})}$ resonances to long-lived ${n(N^{+}=J^{\prime})}$ Rydberg-Stark states. The continuous orange line represents the function $\tau^{*} = 1.34\,n^{1.62} ~\mathrm{\mu s}$ fit to the data.}
    \label{Fig4}
\end{figure}

Deceleration from $v_{\mathrm{i}}=800~\mathrm{m \, s^{-1}}$ to rest in the laboratory-fixed frame of reference, with $a_{\mathrm{t}} = {-3.19 \times 10^{6}~\mathrm{m \, s^{-2}}}$, allowed electrostatic trapping at times $\geq 250~\mathrm{\mu s}$ after excitation. Under these conditions, the decay of the molecules from the trap was monitored through the change in the PFI signal with trapping time, $t_{\mathrm{trap}}$, as shown in Fig.~\ref{Fig4}. It was possible to trap molecules excited on the ${43\mathrm{f}(N^{+}=J^{\prime})}$ resonance for up to 10~ms. However, it can be seen that the trap decay rate reduces over time. For the typical number densities of $\sim 10^{4}~\mathrm{cm^{-3}}$ in the trap, with tens of molecules trapped in each experimental cycle ($\sim5\%$ of those initially prepared in long-lived Rydberg states), collisions do not contribute to this change in decay rate, nor do effects of electric field ionization~\cite{Rayment2022}. The changes observed are therefore attributed to effects of blackbody induced transitions, which become important on longer trapping timescales, and the distribution of Rydberg states initially populated having a range of lifetimes, such that the larger fraction ($>0.9$) of molecules in shorter-lived states decay quickly leaving only molecules in longer-lived states trapped at later times~\cite{Seiler2016}. Expanded views of the signal at early trapping times are shown in Fig.~\ref{Fig4}(b) following excitation on the ${43\mathrm{f}(N^{+}=J^{\prime})}$ (filled blue circles) and ${39\mathrm{f}(N^{+}=J^{\prime})}$ (open orange squares) resonances. To determine effective trap decay time constants, $\tau^*$, from these data, single exponential functions were fit for values of $t_{\mathrm{trap}}=50 -- 500~\mathrm{\mu s}$. The fit was restricted to this time range to avoid spurious effects from the motion of the molecules in the trap at early times~\cite{Zhelyazkova2019,Deller2020PRL}, and the deviation from a single exponential function at late times. The data in Fig.~\ref{Fig4}(b), for molecules excited on the ${43\mathrm{f}(N^{+}=J^{\prime})}$ and ${39\mathrm{f}(N^{+}=J^{\prime})}$ resonances, yielded values of $\tau^*$ of $622\pm26~\mathrm{\mu s}$ and $485\pm 37~\mathrm{\mu s}$, respectively.

The range of Rydberg states suitable for trapping was determined by recording laser photoexcitation spectra with detection at $t_{\mathrm{trap}} = 100~\mathrm{\mu s}$ [Fig.~\ref{Fig2}(b)]. The relative intensities of the resonances in this spectrum reflect (i)~the efficiency with which the long-lived Rydberg-Stark states were populated, (ii)~the deceleration and trap loading efficiencies which for lower (higher) values of $n$ are limited by the ${\sim n^{2}}$ dependence of the maximal electric dipole moment (${\sim n^{-4}}$ dependence of the ionization electric field) of the Rydberg-Stark states, and (iii)~the decay of the Rydberg states. As in Fig.~\ref{Fig2}(a), transitions to states with ${n\mathrm{f}(N^{+}=J^{\prime})}$ character are identified for $\upsilon_{2} < 26780~\mathrm{cm^{-1}}$. However, at higher wave-numbers the high density of transitions to long-lived Rydberg states with $N^{+} = J^{\prime}$ and $J^{\prime}\pm 2$, makes it difficult to assign individual features. The intense resonances at $\upsilon_{2} = 26755.55$ and $26786.38~\mathrm{cm^{-1}}$ correspond to transitions to states with ${40\mathrm{p}(N^{+}=J^{\prime})}$ and ${53\mathrm{p}(N^{+}=J^{\prime})}$ character, respectively. On these resonances, the higher intensities result from effects of charge-quadrupole interactions that mix the accidentally near degenerate ${40\mathrm{p}(N^{+}=2)}$ [${53\mathrm{p}(N^{+}=1)}$] and ${43\mathrm{f}(N^{+}=0)}$ [${43\mathrm{f}(N^{+}=3)}$] states, which are separated by $\sim 0.06~\mathrm{cm^{-1}}$ [$\sim 0.11~\mathrm{cm^{-1}}$], and allow efficient population transfer into long-lived Rydberg-Stark states.

For N$_2$ molecules excited on ${n\mathrm{f}(N^{+} = J^{\prime})}$ resonances to ${n(N^{+}=J^{\prime})}$ Rydberg-Stark states, $\tau^*$ was determined for values of $n$ between 39 and 48 as  in Fig.~\ref{Fig4}(c). These decay times -- each the mean of up to 11 separate measurements -- range from $\sim 450$ to $700~\mathrm{\mu s}$, and increase with $n$. From a least-squares fit, it was found that ${\tau^* = 1.34\,n^{1.62 \pm 0.05}~\mathrm{\mu s}}$. This deviates from the $\sim n^{4}$-dependence of the lifetimes expected for individual $\ell$-mixed Rydberg-Stark states~\cite{Chupka1993}. However, this can be interpreted by direct comparison with measurements of $\tau^*$ for He atoms in Rydberg-Stark states populated by the same mechanism as the Stark states in N$_2$. These data (see supplemental material~\cite{SupplementalMaterial}) show that when samples in Rydberg-Stark states with a range of values of $m_{\ell}$ (or $M_N$ and $N^+$ in the case of N$_2$) are decelerated and trapped, $\tau^*$ increases more gradually with $n$ than if sublevels with a single value of $m_{\ell}$ are populated.

Finally, the increase in $\tau^*$ with $n$ in Fig.\ref{Fig4}(c) contrasts with the decrease seen in comparable measurements with NO~\cite{Deller2020PRL,Rayment2021}. The behavior in NO was attributed to effects of vibrational channel interactions that originate from the interaction of the Rydberg electron with the multipole moments of the NO$^+$ cation. These intramolecular interactions are different in NO and N$_2$ because of the different multipole moments of NO$^+$ and N$_2^+$. This, together with differences in energy-level structure, means that the values of $\tau^*$ for the Rydberg N$_2$ molecules that are most efficiently decelerated and trapped are not so strongly affected by channel interactions with short-lived states that undergo fast nonradiative decay. Since the values of $\tau^*$ in Fig.~\ref{Fig4}(c) are on the order of the fluorescence lifetimes of hydrogenic Rydberg-Stark states with $m_{\ell}=3$ or~4 ($600$ to $1400~\mu$s, or $800$ to $1900~\mu$s for $n = 39 - 48$), it is inferred that the decay of the Rydberg states of the trapped N$_2$ molecules is dominated by spontaneous emission. 

In conclusion, we have reported the first demonstration of electrostatic trapping neutral N$_2$ molecules. This was achieved by excitation from the $\mathrm{X}\,\mathrm{^{1}\Sigma^{+}_{g}}$ ground state to high Rydberg states using a $(2+1')$ resonance-enhanced three-photon excitation scheme and the methods of Rydberg-Stark deceleration. The molecules were trapped in a cryogenically cooled apparatus for up to 10~ms, with effective trap decay time constants $>700~\mu$s. The quantum-state dependence of the decay of the trapped molecules was interpreted by comparison with results of similar measurements with He. The long decay times of the trapped N$_2$ molecules, together with the general increase in these times with $n$ indicate that the decay of the excited Rydberg states was dominated by spontaneous emission rather than predissociation.

In the future, it will be of interest to refine these experiments to allow full rotational and Stark-state selectivity by photoexcitation through the S-branch of the ${\mathrm{X}(J^{\prime\prime}) \rightarrow \mathrm{a^{\prime\prime}}(J^{\prime})}$ transition, using narrower-bandwidth and higher intensity lasers. With such Fourier-transform-limited laser radiation, thousands of rotationally and vibrationally state-selected Rydberg molecules could be trapped in each experimental cycle, at number densities $>10^6$~cm$^{-3}$. These molecules could ultimately be coherently transferred to a selected rotational and vibrational level in the ground electronic state after deceleration to prepare quantum-state-selected, velocity-controlled ground-state samples of N$_2$ with similar number densities for collisions, spectroscopic studies, and loading into optical dipole traps~\cite{singh23a}. More refined measurements of the decay rates of individual Rydberg-Stark states in N$_2$ will also permit direct quantitative comparison of the experimental data with multichannel defect theory calculations.

\begin{acknowledgments}
This work was supported by the European Research Council (ERC) under the European Union’s Horizon 2020 research and innovation program (Grant No. 683341). M. H. R. is grateful to the Engineering and Physical Sciences Research Council (EPSRC) for support through their Doctoral Training Partnership (Grant No. EP/R513143/1).
\end{acknowledgments}

\bibliography{References}

\end{document}